\documentclass[seceq]{ptptex}

\usepackage[dvips]{graphicx}
\usepackage{amsmath}
\usepackage{wrapfig}
\usepackage{feynmp}

\newcommand{\CO}{{\cal O}}

\def\be{\bar{e}}
\def\bu{\bar{u}}
\def\bd{\bar{d}}

\title{On the realization of the MSSM inflation}

\author{Kohei Kamada$^1,^2$ and J. Yokoyama$^2,^3$} 

\inst{$^1$Department of Physics, Graduate School of Science,  
The University of Tokyo, Tokyo 113-0033, Japan\\
$^2$Research Center for the Early Universe (RESCEU),
Graduate School of Science, The University of Tokyo, Tokyo 113-0033, Japan\\
$^3$Institute for the Physics and Mathematics of the Universe,\\
The University of Tokyo, Chiba 277-8582, Japan
}

\abst{
We study the fine-tuning problem on the initial condition of  inflation in the minimal supersymmetric standard model (MSSM), associated with the narrowness of the slow-roll region. 
We consider two cases before the onset of the MSSM inflation, namely,  the radiation dominated era and preceding inflationary era. 
We find thermal dissipation does not induce additional friction to the inflaton, so that it does not help enlarge the slow-roll region. 
We argue thermal correction to the potential in the preceding radiation dominated era and quantum fluctuations in the preceding inflationary era disturb field configuration and prevent the MSSM inflation unless some specific preinflation history is realized. 
}

\begin{document}

\maketitle
\section{Introduction}
The inflationary expansion in the early Universe \cite{Sato:1980yn} is an indispensable ingredient of the modern cosmology not only to solve the flatness and the horizon problems but also to account for the origin of primordial fluctuations \cite{fluctuation}. 
To construct a model of inflation it is reasonable to respect supersymmetry (SUSY) \cite{Nilles:1983ge}, because in supersymmetric theories radiative corrections which tend to break the flatness of the potential required for inflation are suppressed.  

For the same reason, a singlet with respect to the gauge group of the Standard Model (SM) of particle physics is usually introduced to realize inflation.  
Recently, however, Allahverdi et al. \cite{Allahverdi:2006iq,Allahverdi:2006we} proposed a model of inflation based on the supersymmetric standard model (MSSM).  
In this model, the inflaton is identified with a flat direction in the MSSM \cite{Gherghetta:1995dv}, a gauge invariant combination of either squark or slepton fields.
Flat directions are lifted by SUSY-breaking effects and a non-renormalizable potential. 
If model parameters are fine-tuned, the potential along the flat direction may have a saddle point and inflation can occur there. 
In this MSSM inflation model, the slow-roll phase is driven by the third derivative of the potential. 
If we take the inflaton as the $\bu \bd\bd$ or the $LL\be$ flat direction, which we denote by $\varphi$, the saddle point is located at $ \varphi \sim 10^{14-15}$ GeV and the Hubble parameter during inflation is about $10^{-2}-1$ GeV. 
In this case we have the right amplitude of primordial fluctuations. 
The most interesting feature of the MSSM inflation is that the inflaton couplings to particles in the SM are known and, at least in principle, measurable in laboratory experiments such as the Large Hadron Collider or a future Linear Collider. 
Moreover, if the MSSM inflation takes place, this inflation would be the last inflation that we observe. 
Therefore if its number of e-folds is sufficiently large, the information before the MSSM inflation is washed out and we cannot acquire any information on the prior epoch. 
On the contrary, if it is small, the MSSM inflation can help models of high-energy inflation that has problems such as the smallness of the  number of e-folds or the lack of reheating, which are the problems of many string inflation models \cite{McAllister:2007bg}. 

Despite these attractive features described above, the MSSM inflation has some drawbacks such as the fine-tuning of the model parameters \cite{BuenoSanchez:2006xk} and worse, the fine-tuning of the initial condition. 
In particular, the latter is a very difficult problem. 
Because the slow-roll region for the MSSM inflation is extremely narrow, if the former is solved by some symmetry, the inflaton must set on the saddle point highly accurately in order to expand the Universe exponentially.  

Some solutions have been proposed by the advocates of this model \cite{Allahverdi:2007wh,Allahverdi:2008bt} in the context of false vacuum inflations before the MSSM inflation. 
They claim that the inflaton candidate enters the slow-roll region for the MSSM inflation during the preceding inflation with a very small velocity.  
This is due to an attractor behavior near the inflection point during the quasi-de Sitter expansion of the Universe. 
This solves not only the initial condition problem of the MSSM inflation but also the graceful exit problem of a false vacuum inflation. 
The key point of the mechanism above is the enhancement of the friction term in the equation of motion. 

There is another possible mechanism that may enhance the friction term. 
Interactions between an inflaton and other fields can cause dissipative effects on the dynamics. 
This phenomenon has been discussed in the high temperature regime \cite{Hosoya:1983ke,Morikawa:1984dz,Morikawa:1986rp,Gleiser:1993ea,Yokoyama:2004pf,Yokoyama:1998ju}, in particular, in the context of the warm inflation \cite{Berera:1995ie}. 
Moreover, recently it was suggested that even in low-temperature regime, there can arise dissipative effects via the interactions catalyzed by heavy fields \cite{Moss:2006gt,Berera:2008ar,Graham:2008vu}. 
Therefore it is worthwhile to investigate whether a thermal effect can stop the inflaton near the inflection point during radiation dominated era or the period when thermal plasma exists even though it is a subdominant component of the Universe. 
In particular, if the MSSM inflation had a thermal friction dominated inflationary solution, the prediction about the MSSM inflation would change dramatically. 

On the other hand, there is an unanswered question about the previous solution to the initial condition tuning. 
In the previous study, the authors have paid less attention to the effect of stochastic behaviors on the scalar fields in the quasi-de Sitter background. 
The stochastic behavior in the quasi-de Sitter Universe is common not only to the inflaton but also to any scalar fields in the quasi-de Sitter background \cite{Stoch,Starobinsky:1994bd}. 
Scalar fields in the quasi-de Sitter background feel a stochastic force due to the horizon crossing of short-wavelength quantum fluctuations. 
The distribution of those scalar fields becomes rather probabilistic, and the analysis of the classical equation alone would be insufficient. 

The aim of this paper is to clarify how  the dissipative effect and the stochastic behavior affect the initial condition for the MSSM inflation. 
The structure of the paper is as follows. 
In the following section, we present the relevant part of the MSSM in which we use. 
Then we consider the fine-tuning problem of the MSSM inflation. 
In section \ref{secDis}, we consider the probability of the emergence of the MSSM inflationary epoch after the radiation dominated era. 
We discuss how dissipative effects affect the dynamics of inflaton. 
Then we show that regrettably they do not help the tuning of the initial condition of the MSSM inflation. 
We argue that there are no thermal friction-dominated inflationary solutions \cite{Berera:1995ie} on the MSSM inflation and it is difficult to stop the MSSM inflaton near the inflection point with a sufficiently small velocity.
In section \ref{secSto} we discuss the stochastic behavior on the MSSM inflaton in the de Sitter background. 
We show that stochastic force likely disturbs the initial condition tuning during the preceding inflation. 
Finally, section \ref{conc} is devoted to the conclusion. 

\section{MSSM inflation \label{sec2}}

Let us summarize the main features of the MSSM inflation \cite{Allahverdi:2006iq} briefly. 
We add a non-renormalizable superpotential of the form \cite{Gherghetta:1995dv} 
\begin{equation}
W_{\rm non}=\frac{\lambda}{6M_G^3}\Phi^6 \label{nonsup}
\end{equation} 
to the superpotential of MSSM. 
Here $M_G$ is the reduced Planck scale and $\Phi$ is a supermultiplet whose scalar component $\phi$ parameterizes the flat direction to play the role of this inflaton. 
In the case where the $\bu \bd \bd$ flat direction acts as an inflaton, 
$\phi$ represents the following field configuration.  
\begin{equation}
{\tilde \bu}_i^\alpha={\tilde \bd}_j^\beta ={\tilde \bd}_k^\gamma=\frac{1}{\sqrt 3}\phi, \ \ (j\not=k,\alpha\not=\beta\not=\gamma), 
\end{equation}  
where $i,j$ and $k$ are the family indices and $\alpha,\beta$ and $\gamma$ are the color indices.
We assume that $\lambda$ is a parameter of order unity. 
The flat direction is lifted by the superpotential \eqref{nonsup}. 
Hereafter we consider the case when the scalar component of $\phi$ acquires a large expectation value. 

Including the SUSY breaking effect from the hidden sector, 
the scalar potential is found to be
\begin{equation}
V(\phi)=m_\phi^2|\phi|^2+\left(\frac{a \lambda}{6M_G^3}\phi^6+{\rm H.c.}\right)+\frac{\lambda^2}{M_G^6}|\phi|^{10}. 
\end{equation}
Here $m_\phi \sim $100 GeV -- 1 TeV is the soft SUSY-breaking mass of $\phi$ and $a$ is a complex parameter whose amplitude is also $|a|\simeq 100$ GeV -- $1$ TeV. 
This applies when SUSY-breaking is gravity-mediated. 

After minimizing the potential along the angular direction of $\phi$, and rewriting the flat direction as $\phi=\dfrac{1}{\sqrt 2}\varphi e^{i \theta_{\rm min}}$,  with $\varphi$ being the real amplitude, 
the scalar potential is found to be
\begin{equation}
V(\varphi)=\frac{1}{2}m_\phi^2 \varphi^2-\frac{A \lambda}{24 M_G^3}\varphi^6+\frac{\lambda^2}{32M_G^6}\varphi^{10}. 
\end{equation}
Here $\theta_{\rm min}$ is a value that satisfies $\cos(6 \theta_{\rm min}+\theta_a)=-1$ where $ \theta_a$ is the phase of $a$ and we have defined $A\equiv |a|$. 

If $A$ is fine-tuned as $A^2= 20m_\phi^2(1+\alpha^2/4)$ with $\alpha^2 \ll 1$,  $V(\varphi)$ has a inflection point $\varphi_0$ at
\begin{align}
&\varphi_0 = \left(\frac{2AM_G^3}{5\lambda}\right)^{1/4}(1+\CO(\alpha^2))=4.0\times 10^{14} \lambda^{-1/4}\left(\frac{m_\phi}{10^3{\rm GeV}}\right)^{1/4}(1+\CO(\alpha^2)){\rm GeV}, \\
&V^{\prime \prime}(\varphi_0) = 0, 
\end{align} 
with
\begin{align}
&V(\varphi_0)= \frac{4}{15}m_\phi^2\varphi_0^2(1+\CO(\alpha^2)), \\
&V^{\prime}(\varphi_0) = m_\phi^2\varphi_0 \times \CO(\alpha^2).  
\end{align}
Therefore inflation can occur near the inflection point and the Hubble parameter takes
\begin{equation}
H_{\rm MSSM}\simeq \frac{2m_\phi}{3{\sqrt 5} M_G}\varphi_0 \simeq 4.9\times 10^{-2}  \left(\frac{m_\phi}{10^3 \rm{GeV}}\right)\left(\frac{\varphi_0}{4.0\times 10^{14}{\rm GeV}}\right){\rm GeV},
\end{equation}
there. 
Neglecting $\alpha$,  the potential for MSSM inflation $V(\varphi)$ can be expanded around the inflection point as 
\begin{equation}
V(\varphi)=\frac{4}{15}m_\phi^2\varphi_0^2+\frac{16}{3}\frac{m_\phi^2}{\varphi_0} (\varphi-\varphi_0)^3. \label{plateau}
\end{equation}

Now we comment on the slow-roll condition and the number of e-folds. 
The slow-roll parameters are defined as
\begin{align}
\epsilon & \equiv \frac{1}{2}M_G^2\left(\frac{V_{,\varphi}}{V}\right)^2, & \eta&\equiv M_G^2\left(\frac{V_{,\varphi\varphi}}{V}\right). 
\end{align}
Here the subscript $_{, \varphi}$ represents the derivative with respect to $\varphi$. 
The slow-roll conditions $\epsilon,|\eta| \ll 1$ are written as 
\begin{equation}
\frac{\Delta \varphi}{\varphi_0}\ll\frac{\varphi^2_0}{120M_G^2}=9.0\times10^4 \left(\frac{\varphi_0}{4.0\times 10^{14} {\rm GeV}}\right)^2{\rm GeV}, \ \   \Delta \varphi \equiv |\varphi-\varphi_0|. 
\end{equation} 
Using the slow-roll equation
\begin{equation}
3H{\dot \varphi}+V_{, \varphi}(\varphi)=0, 
\end{equation}
the number of e-folds ${\cal N}$ during the period when the evolution of the inflaton determined by the classical slow-roll regime can be calculated as 
\begin{align}
{\cal N}&=\int_{t_i}^{t_{\rm end}} H dt=\int_{\varphi_i}^{\varphi_{\rm end}} \frac{H}{\dot \varphi}d\varphi \notag \\
&=\frac{\varphi^3_0}{60 M_G^2(\varphi_0-\varphi_i)}=\frac{5^{3/4}\pi^{1/2}}{10}\frac{\varphi_0}{M_G^{1/2}m^{1/2}}. 
\end{align}
Here $\varphi_i=\left(1+\dfrac{\varphi_0m^{1/2}}{6\times5^{2/3}\pi^{1/2}M_G^{3/2}}\right)\varphi_0$ is the point where the classical force gets stronger than quantum fluctuation, $H_{\rm MSSM}/2\pi<{\dot \varphi}/H_{\rm MSSM}$ and $\varphi_{\rm end}=\varphi_0-\varphi_0^3/(120M_G^2)$. 

The amplitude of the curvature perturbations on the scale which left the Hubble radius at ${\cal N}_{\rm COBE}$ e-folds before the end of inflation is given by  
\begin{equation}
\Delta_{\cal R}= \frac{H^2}{2 \pi{\dot \varphi}} \simeq 10^{-4}\times \left(\frac{m_\phi}{10^3 {\rm GeV}}\right)\left(\frac{4.0\times 10^{14}{\rm GeV}}{\varphi_0}\right)^2\left(\frac{{\cal N}_{\rm COBE}}{50}\right)^2 
\end{equation}
The spectral index of the power spectrum $n_s$ is given by  
\begin{equation}
n_s=1+2\eta-6\epsilon \simeq 1-\frac{4}{{\cal N}_{\rm COBE}}. 
\end{equation}
The typical value of ${\cal N}_{\rm COBE}$ is 50. 
So this value is rather small as compared to the observational data \cite{Komatsu:2008hk}. 
While this model has not been observationally ruled out yet, this feature may be regarded as a drawback. 
Planck \cite{Planck} or other future experiments will provide us with further information on it. 
On the other hand, if the number of e-folds of MSSM inflation turns out to be shorter, we would be free from such large-scale constraints. 
But this would require even more stringent fine tuning as we see later.

Next we incorporate effects of nonvanishing $\alpha$ \cite{Allahverdi:2006iq}. 
When $\alpha^2>0$, there are two potential extrema, 
\begin{equation}
\varphi_\pm=\varphi_0\left(1\pm\frac{\alpha}{2}\right) .
\end{equation}
We find two classes of inflationary solutions with different  dynamics of the flat direction. 
One is the case the field starts slow-roll evolution at a smaller value than the local maximum of the potential. 
The other is that the field is once trapped in the false vacuum and tunnel to its vicinity out of it. 
The first solution requires the slow-roll parameter $\eta$ to be small at the local maximum of the potential, that is, 
\begin{equation}
|\eta|\ll 1 \Leftrightarrow \alpha \ll \frac{\varphi_0^2}{60M_G^2}\sim 10^{-11}\label{alp_con}.
\end{equation}
The conditions for the second solution, on the other hand, are less stringent. 
However, the slow-roll phase must take place in the dynamics of the flat direction in order to generate right primordial perturbations after tunneling, which requires the same constraint on $\alpha$  as \eqref{alp_con}. 

When $\alpha^2 < 0$, on the other hand, there is no extremum. 
In this case, the condition to realize inflation is to acquire the sufficient number of e-folds and to generate the right amplitude of  primordial density fluctuations. 
This condition can be expressed as
\begin{equation}
|\alpha| \lesssim 10^{-9}-10^{-8}.
\end{equation}
In both cases, as a conclusion, the relation between $A$ and $m_\phi$ requires a high degree of fine-tuning. 

We now turn to the problem of the initial condition. 
Even if the fine-tuning of the model parameters described above is realized by some symmetry, the MSSM inflation suffers from the severe initial value problem. 
In order to see this, hereafter we set exactly $\alpha=0$.  

If we set the time derivative of the inflaton ${\dot \varphi}_+$ (at $t=t_+$) at the upper boundary of the slow-roll region, 
$\varphi=\varphi_0+\varphi_0^3/(120M_G^2) \equiv \varphi_{\rm c+}$, 
its subsequent time evolution is given by
${\dot \varphi}={\dot \varphi}_+e^{-3H(t-t_+)}$, 
neglecting the potential force and assuming that the Universe is dominated by the energy density of $\varphi$. 
If the inequality
\begin{equation}
H^{-1}\left.\frac{d \log \varphi}{dt}\right|_{\varphi=\varphi_{\rm c+}} > \frac{\varphi_0^2}{20M_G^2},
\end{equation} 
holds, the inflaton passes through the slow-roll region within a time interval of $H^{-1}$ before being decelerated. 
As a consequence, inflation does not occur. 
Given that the slow-roll region is very narrow, $\Delta \varphi \simeq \varphi_0^3/60M_G^2 \simeq 10^{-10}\varphi_0$, 
we conclude that the inflaton must reach the slow-roll region with an extremely small velocity. 

Suppose that the time derivative of the inflaton vanishes at some field value $\varphi_i$ above the inflection point, ${\dot \varphi}=0|_{\varphi=\varphi_i >\varphi_0}$, at the beginning of our Universe or during the oscillation of the inflaton around the origin. 
Then, the friction term is negligible when ${\dot \varphi}$ is small enough and the inflaton obeys the equation, 
\begin{equation}
\frac{1}{2}{\dot \varphi}^2(\varphi=\varphi_{\rm c+})\simeq V(\varphi_i)-V(\varphi_{\rm c+}). 
\end{equation}
Therefore ${\dot \varphi}$ is larger than the threshold value if $\varphi_i > \varphi_0^3/(10 M_G^2)\equiv \varphi_{i{\rm c}}$. 
By the numerical analysis we have confirmed that the situation does not change considerably even if we consider the effect of the Hubble friction correctly. 
This means that the inflaton must stop once in the tiny range just a little wider than the slow-roll region. 

If the MSSM inflaton starts its evolution at a value larger than the inflection point without fine-tuning, it oscillates around the potential minimum. 
The amplitude of its oscillation decreases due to the cosmic expansion and the particle production. 
When the amplitude approaches the inflection point, the dynamics of the oscillation is almost that of harmonic oscillator and can be written as 
\begin{align}
\varphi&\simeq\frac{\varphi_I}{m_\phi t}\sin (m_\phi t), \\
{\dot \varphi}&\simeq\frac{\varphi_I}{t}\cos (m_\phi t),
\end{align}
where $\varphi_I \lesssim M_G$ is the amplitude at the onset of oscillation. 
Here we have only considered the effect of the cosmic expansion. 
Taking $\varphi_I\gg \varphi_0$, the energy of the inflaton decreases by $(2\pi \varphi_0/\varphi_I)V(\varphi_0)$ after one oscillation when $\varphi \simeq \varphi_0$. 
Therefore the inflaton loses its energy by at least $10^{-3}V(\varphi_0)$ in this period. 
The amplitude of the oscillation decreases by $\Delta \varphi\simeq 10^{-1}\varphi_0$ near the inflection point after one oscillation, 
that is, 
\begin{equation}
\Delta V|_{\varphi\simeq\varphi_0}=\frac{16}{3}\frac{m_\phi^2}{\varphi_0}\Delta \varphi^3. 
\end{equation}
Using the full potential, we have numerically confirmed the amplitude of the oscillation decreases by $5\times 10^{-2}\varphi_0$ near the inflection point. 
This range is much larger than the range estimated above, $\varphi_{ic}-\varphi_0=\varphi_0^3/(10 M_G^2)\simeq 10^{-10}\varphi_0$. 
Therefore we conclude that the extremely severe fine-tuning of the initial condition of order of $\CO(10^{-9})$ is needed for the MSSM inflation to take place after the inflaton oscillation dominated era.

\section{Dissipative effect\label{secDis}}

In the previous section we have reviewed the features of the MSSM inflation. 
In particular, we have seen that the fine-tuning problem of the potential parameters and the initial conditions are extremely severe. 
This is due to the narrowness of the slow-roll region. 

In the previous argument, we set the time when the MSSM inflaton comes to overwhelm the energy density of the Universe as the initial point. 
Here we consider the dynamics of $\varphi$ in the case the prior stage was radiation dominant. 

If the cosmic temperature $T$ was higher than $h\varphi$, where $h$ is the smallest Yukawa coupling of $\varphi$
\footnote{Here we take $h\lesssim 10^{-3}$ because the $\bu\bd\bd$ flat direction contains the first or the second family.}, then it would acquire a thermal mass and rolls down towards $\varphi=0$. MSSM inflation cannot occur with such an initial condition. 
For $T<h\varphi$, on the other hand, the potential of the MSSM inflaton does not suffer from thermal corrections, but still its dynamics may be affected by the presence of subdominant thermal fields. 
As has been studied in the literatures \cite{Hosoya:1983ke, Morikawa:1984dz, Morikawa:1986rp, Gleiser:1993ea}, in particular in the context of the warm inflation \cite{Berera:1995ie}, they might cause dissipative effects and modify the dynamics of the inflaton although it is not easy in general \cite{Yokoyama:1998ju}. 
As a consequence, the narrowness of the slow-roll region could be relaxed if dissipative effects turned out to be strong enough. 
It is nontrivial whether the MSSM inflation takes place or not in this situation. 
In this section, we study whether it can relax the problems of the MSSM inflation or not.

\subsection{modification to inflation}

First we incorporate a thermal bath as a subdominant component of the Universe during the inflaton dominated era. 
As the inflaton $\varphi$ has interaction with other fields a dissipative phenomenon takes place and the inflaton feels a damping force, in particular, when they are in the thermal bath. 
In this case there exists energy transfer from the inflaton to the radiation and the latter is not diluted away completely. 
Recall, however, that all the fields must be heavy which are directly coupled to $\varphi$ in order to avoid thermal corrections to the potential. 
We therefore consider the case the inflaton couples to the radiation catalyzed by heavy fields as considered in \cite{Moss:2006gt,Berera:2008ar,Graham:2008vu}. 
In such a case, we do not have to worry about the thermal correction to the potential. 

According to the discussion above, we take into account dissipative effects and write down the equation of motion for the inflaton $\phi$ approximately as \cite{Morikawa:1984dz,Albrecht:1982mp,Yokoyama:1987an}
\begin{equation}
{\ddot \phi}+(3H+F_r){\dot \phi}+\frac{\partial V}{\partial \phi}=0. \label{EOM}
\end{equation}
Here $F_r$ is the dissipative coefficient representing the dissipative effect and depends on $\phi$ and the temperature of radiation $T$. 
We estimate the value of $F_r$ in the next subsection. 
The relative strength of the dissipative effect compared to the friction term from the expansion is denoted by, 
\begin{equation}
r\equiv\frac{F_r}{H}. 
\end{equation}
If $r$ is much larger than unity, the dynamics of inflaton would be modified. 
In order to estimate it quantitatively, we evaluate the dynamics of the inflaton and other components of the Universe further. 

Other equations that governs the universe are, 
\begin{align}
3M_G^2H^2&=V(\phi)+\rho_\gamma, \label{Frie}\\
{\dot \rho}_\gamma+4H\rho_\gamma&=F_r{\dot \phi}^2,\label{Boltz}
\end{align}
where $\rho_\gamma=(\pi^2/30)g_*T^4$ is the energy density of the radiation,
with $g_*$ being the effective number of relativistic degree of freedom.  

The slow-roll conditions in this case are the conditions that time variation of $H$, ${\dot \phi}$ and $T$ is negligible in comparison to the time scale of the cosmic expansion \cite{Hall:2003zp}. 
Using equations \eqref{EOM}, \eqref{Frie} and \eqref{Boltz}, if $r \gg 1$, the slow-roll condition is modified as \cite{Hall:2003zp}
\begin{equation}
\epsilon \ll r,  \ \ \ \eta \ll r,  \ \ \ \beta \ll r. \label{newsl}
\end{equation}
Here $\beta$ is the new slow-roll parameter introduced in order to take into account the time variation of $F_r$, 
\begin{equation}
\beta \equiv M_G^2 \left(\frac{V_{,\phi} F_{r,\phi}}{V F_r}\right). 
\end{equation} 
We can see from \eqref{newsl} that slow-roll condition is relaxed by the factor of $r$. 
\footnote{Strictly speaking, the degree of relaxation is dependent on the form of the potential $V$ and the dissipative coefficient $F_r$. }

\subsection{dissipative coefficients}

As discussed above, we consider the case where the inflaton couples to the radiation catalyzed by heavy fields. 
Let us consider the superpotential of the form, 
\begin{equation}
W=h_1\Phi X^2+h_2 XY^2,  
\end{equation}
in addition to \eqref{nonsup}. 
Here $h_1$ and $h_2$ are the Yukawa couplings, $\Phi$ is the inflaton, $X$ is an intermediate heavy field, and $Y$ is the light field that is in thermal bath. 
If we consider the case where the $\bu \bd \bd$ flat direction acts as the inflaton, $X$ represents the left-handed quark multiplet and the Higgs multiplet 
and $Y$ represents the left and right-handed lepton multiplet. 
The relevant part of the scalar potential are 
\begin{equation}
V_{\rm int}=4h_1^2 |\phi|^2|\chi|^2+h_2^2 |y|^4 + 2h_1 h_2(\phi\chi y^{*2}+\phi^*\chi^*y^2)+4h_2^2|\chi|^2|y|^2 \label{potint}
\end{equation}
Here $\phi, \chi, y$ are the scalar components of $\Phi, X,Y$ respectively. 
From \eqref{potint}, we can see that $\chi$ acquires a large mass, $m_\chi=h_1 |\phi_c|$, where $\phi_c$ is the value of the inflaton. 

The dissipative coefficient arising from this interaction in low temperature regime $T \ll h_1 \phi$ is well approximated by \cite{Moss:2006gt,Berera:2008ar}
\begin{equation}
F_r\simeq 0.16h_2^4{\cal N}_d\frac{T^3}{\phi^2} \equiv C_\phi\frac{T^3}{\phi^2}.\label{fricth} 
\end{equation}
Here ${\cal N}_d$ is the number of the dissipative channels with the same coupling strength $h_2$. 
This is due to the particle production of $y$ fields by the motion of $\phi$ and the interaction via the heavy field $\chi$. 
This effect is suppressed by a factor of $T^2/m_\chi^2$, therefore we have $T^3$ dependence. 

As a consequence, \eqref{EOM} reads 
\begin{equation}
{\ddot \phi}+3H{\dot \phi}+C_\phi \frac{T^3}{\phi_c^2}{\dot \phi}+\frac{\partial V}{\partial \phi}=0. 
\end{equation}
In deriving \eqref{fricth}, we have neglected the effect of the cosmic expansion and assumed time variation of inflaton to be very small. 
Therefore the time scale of the dissipative process must be much smaller than that of the cosmic expansion and the inflaton motion. 
In other words, this estimate is valid only when $H\ll \Gamma_y(T)$ and $|{\dot \phi}|/\phi\ll \Gamma_y(T)$. 
Here $\Gamma_y(T)$ is the decay width of $y$.

\subsection{application to MSSM inflation}

In this subsection we consider if the dissipative process may affect the dynamics of the MSSM inflation with the following properties: 
\begin{align}
\varphi_0&=\left(\frac{2AM_G^3}{5\lambda}\right)^{1/4}\simeq 4.0\times 10^{14} {\rm GeV}, \label{vev}\\
H_{\rm MSSM}&=\frac{2m_\phi}{3{\sqrt 5} M_G}\varphi_0 \simeq 10^{-16} \left(\frac{m_\phi}{10^3 \rm{GeV}}\right)\varphi_0 \simeq 10^{-1} {\rm GeV}, \label{Hubble}\\
F_r&=0.16h_2^4{\cal N}_d \frac{T^3}{\varphi_0^2} \equiv C_\phi \frac{T^3}{\varphi_0^2}. \label{diss}
\end{align} 
Assuming $H_{\rm MSSM} \ll F_r$, we consider the following quasi-static slow-roll equations, 
\begin{align}
F_r{\dot \varphi}&=8\frac{m_\phi^2}{\varphi_0}(\varphi-\varphi_0)^2, \label{sl1}\\
4H_{\rm MSSM}\rho_\gamma&=F_r {\dot \varphi}^2, \label{sl3}
\end{align}with
\begin{equation}
\rho_\gamma = \frac{\pi^2 g_*}{30}T^4,\label{tempe}
\end{equation}
where $g_*$ is the number of relativistic degrees of freedom and in the context of MSSM $g_*=\CO(10^2)$. 
The only efficient dissipative channel is the process of the production of the third generation of leptons whose Yukawa coupling can be $h_2 \simeq \CO(10^{-1}-1)$ via the heavy Higgs. 
Therefore we can take $C_\phi \simeq \CO(10^{-1})$ at most. 

If we could find a solution to the slow-roll equations \eqref{sl1} and \eqref{sl3} which satisfies the condition
\begin{equation} 
3H_{\rm MSSM}\ll F_r, \ \ \ \rho_\gamma \ll V(\varphi_0), \ \ \ F_r \ll \Gamma_y(t)=\frac{h_2^4 T}{192\pi}, \label{const}
\end{equation}
there would be a friction-dominated inflationary solution to the dynamics of the MSSM inflaton and 
we could conclude that the dissipative effect can relax the slow-roll condition to the MSSM inflation. 

However, the first two conditions of \eqref{const} read, 
\begin{align}
T \gg 6.0 \times 10^{9} {\rm GeV} \lambda^{-1/4} \left(\frac{C_\phi}{10^{-1}}\right)^{-1/3} \left(\frac{m_\phi}{10^3 {\rm GeV}}\right)^{7/12}  \label{Tup}\\
T \ll 7.8\times 10^8 {\rm GeV}  \lambda^{-1/8}\left(\frac{g_*}{10^2}\right)^{-1/4} \left(\frac{m_\phi}{10^3 {\rm GeV}}\right)^{5/8}.  
\end{align}
In order to realize a friction-dominated inflationary solution one must increase $C_\phi$ and/or $\lambda$. 
It is difficult to increase the number of decay channel, $C_\phi$, because it requires new physics below the cut off scale. 
So here we consider only the change of the cut off scale, $M_G/\lambda$. 
However this, too,  does not lead a friction-dominated inflationary MSSM inflation. 
From the above equations \eqref{sl1}, \eqref{sl3} and \eqref{tempe}, we have
\begin{align}
T&=3.2\times 10^{10}\left(\frac{g_*}{10^2}\right)^{-1/7}\left(\frac{C_\phi}{10^{-1}}\right)^{-1/7} \left(\frac{m_\phi}{10^3 {\rm GeV}}\right)^{3/7} \left(\frac{\varphi_0}{10^{15}{\rm GeV}}\right)^{3/7}\left(\frac{|\varphi-\varphi_0|}{\varphi_0}\right)^{4/7}{\rm GeV}.  \label{T-sl}
\end{align}
Because we want to discriminate whether the slow-roll region for the inflaton is enhanced or not, 
we assign $|\varphi -\varphi_0|\simeq \varphi_0(\varphi_0/M_G)^2$. 
Considering $\varphi_0\propto \lambda^{-1/4}$,  from \eqref{T-sl} we have
\begin{equation}
T=1.0\times 10^6 \lambda^{-11/28} \left(\frac{g_*}{10^2}\right)^{-1/7}\left(\frac{C_\phi}{10^{-1}}\right)^{-1/7} \left(\frac{m_\phi}{10^3 {\rm GeV}}\right)^{23/28}{\rm GeV} \label{tt}
\end{equation} 
in this region. 
Because \eqref{tt} cannot satisfy \eqref{Tup} for $\lambda >1$, we cannot have a thermal friction-dominated solution on the MSSM inflation. 
In order for the MSSM inflation to take place, the MSSM inflaton must still reach the very narrow slow-roll region with an extremely small velocity. 

\subsection{Transition from the radiation dominated Universe to the MSSM inflationary regime}
Having seen that thermal friction does not enlarge the slow-roll region, we next examine if the MSSM inflaton can reach this narrow region during the radiation dominated era. 
As discussed above, the cosmic temperature must not be higher than $h \varphi_0 \lesssim 10^{11}\lambda^{-1/4}$ GeV because otherwise $\varphi$ will overshoot the slow-roll region by the thermal correction to the potential. 
So we consider the case where the highest cosmic temperature is less than $h\varphi_0 \lesssim 10^{11}\lambda^{-1/4}$ GeV. 
Moreover, the cosmic temperature at the transition from the radiation dominated era to the MSSM inflation is $T\simeq (m_\phi H_{\rm MSSM})^{1/2} \simeq 10^9\lambda^{-1/8}$ GeV, so the radiation dominated era may last only a few number of e-folds.

In the case  $\varphi \gg \varphi_0$ at the onset of the radiation dominated era, thermal friction term is no larger than the Hubble friction term. 
Then, the square of the Hubble parameter $\pi^2 g_* T^4/ (90 M_G^2)$ is less than the curvature of the potential $3\lambda^2\varphi^8 /M_G^6$. 
So $\varphi$ starts oscillating around the origin and is damped to the origin quickly. 
As a consequence, the MSSM inflation does not take place. 

If $\varphi \gtrsim \varphi_0$ at the onset of the radiation dominated era, $\varphi$ can start moving with a small velocity. 
In this case, however, when $\varphi$ overwhelms radiation, the same situation as discussed in Sec. \ref{sec2} arises because the Hubble friction terms is larger than the thermal friction term at that time. 
So we need the same fine-tuning for the MSSM inflation to take place as the case discussed in Sec. \ref{sec2}. 
In the end, $\varphi$ must be near the slow-roll region for the MSSM inflation at the onset of the radiation dominated era for the MSSM inflation to take place. 

\section{Stochastic approach to initial condition problem \label{secSto}}
In Sec. \ref{sec2}, we have seen that the condition for the MSSM inflaton, 
\begin{equation}
H^{-1}\frac{d \log \varphi}{dt}<\frac{\varphi_0^2}{20 M_G^2}, 
\end{equation}
at $\varphi = \varphi_{c+}$, in order for the MSSM inflation to take place. 
In the previous section we have found this situation does not change even when we consider dissipative effects associated with  the existence of the thermal plasma. 
In other words, the MSSM inflaton must be set at the narrow slow-roll region $(1-10^{-10})\varphi_0<\varphi<(1+10^{-10})\varphi_0$ with negligible velocity and a large nearly homogeneous domain when the MSSM inflaton dominates the cosmic energy density. 
It has been argued in Ref. \cite{Allahverdi:2007wh, Allahverdi:2008bt}, that another inflation before the MSSM inflation, such as false vacuum inflation, may solve the initial value problem. 
In this scenario the MSSM inflaton can approach the inflection point very slowly even if its initial value is far from the inflection point. 
This is because the number of e-folds of the false vacuum inflation can be very large.  
The authors of Ref. \cite{Allahverdi:2007wh, Allahverdi:2008bt}, however,  paid attention to the tuning of the field value but not so much to the effect of the quantum fluctuation in de Sitter space and the resultant spatial inhomogeneity. 
In this section we consider it using the stochastic approach \cite{Stoch, Starobinsky:1994bd} and see whether the preceding inflation can solve the initial value problem including the issue of the spatial homogeneity. 

\subsection{Initial condition tuning by preceding inflation}

It is conceivable that the Universe experiences many stages of inflation at high energy scale. 
Therefore here we consider the case the MSSM inflation follows some other inflations. 
In this case, we must pay attention to the Hubble induced mass. 
When the MSSM inflaton acquires a positive Hubble induced mass, it quickly dumps into the origin and the MSSM inflation does not occur. 
When the MSSM inflaton $\varphi$ acquires a negative Hubble-induced mass term, $-cH^2 \varphi^2$ with $c=\CO(1)>0$, it is driven to the minimum 
\begin{equation}
\varphi_{\rm min}\simeq \left(4\sqrt{\frac{c}{5}}\frac{HM_G^3}{\lambda}\right)^{1/4}
\end{equation}
which is larger than $\varphi_0$ if $H\gg m_\phi/\sqrt{c}$. 
Let us consider the case preceding inflation occurs successively as a false vacuum inflation in a manner like the chain inflation scenario \cite{Freese:2004vs} in stringy landscape. 
Then $\varphi$ decreases according to the false vacuum energy density. 
In particular, when the Hubble parameter due to the preceding inflation sector, $H_{\rm fv}$, satisfies $H_{\rm fv} \ll m_\phi$, we find \cite{Allahverdi:2007wh}
\begin{align}
\varphi_{\rm min} &\simeq \varphi_0\left( 1+\frac{\sqrt{2c}H_{\rm fv}}{4 m_\phi}\right)\\ 
V^{\prime \prime}(\varphi_{\rm min}) &\simeq  8\sqrt{2c} H_{\rm fv} m_\phi
\end{align}
We also find $3 H_{\rm fv}^2M_G^2<V(\varphi_{\rm min})$ for $H_{\rm fv}\lesssim H_{\rm MSSM}$. 
Thereafter the chain-inflation sector is no longer the dominant component of the Universe, and it can no longer govern the evolution of $\varphi$. 
Since $V^{\prime \prime}(\varphi_{\rm min}) \gg H_{\rm fv}^2$ at this time $\varphi$ is located at $\varphi_{\rm min} \simeq \varphi_0[1+\CO(10^{-5})]$, which is still much larger than $\varphi_{c+}$; the critical value for the slow-roll MSSM inflation. 
The dynamics in this regime, in particular the fate of $\varphi$, cannot be solved without specifying the detailed structure of the chain-inflation sector and its interaction with $\varphi$, which is beyond the scope of the present paper.

Next we consider the case without Hubble induced mass \cite{Allahverdi:2007wh, Allahverdi:2008bt}. 
If the Hubble parameter of the preceding inflation is less than $m_\phi$, the slow-roll region of $\varphi$ is limited to a tiny vicinity around the inflection point, $|\varphi-\varphi_0|/\varphi_0 \ll (H/m_\phi)^2$. 
 If the MSSM inflaton starts its evolution in the region where the potential can be approximated as $\lambda^2\varphi^{10}/M_G^6$ at the beginning of the preceding inflation, it oscillates around the potential minimum decreasing its amplitude. 
When the amplitude approaches the inflection point, the dynamics of the oscillation can be written as 
\begin{equation}
\varphi\simeq \varphi_* e^{-3Ht/2}\sin(m_\phi t),  
\end{equation}
where $\varphi_*$ is the amplitude of the MSSM inflaton at $t=0$. 
The amplitude of the oscillation changes by 
\begin{equation}
\Delta \varphi \simeq \left(\frac{9H}{16m_\phi}\right)^{1/3}\varphi_0
\end{equation}
in one oscillation. 
Here we have evaluated $\Delta \varphi$ as was done in Sec.\ref{sec2}. 
Because $H/m_\phi>H_{\rm MSSM}/m_\phi \simeq 10^{-4}$, the required fine-tuning for the initial value in order to locate $\varphi$ in the narrow slow-roll range after some oscillation is as severe as the case in Sec.\ref{sec2}.  

If $\varphi$ is set around the tiny slow-roll region, however, it can still be there even at the end of the preceding inflation and eventually enters the slow-roll region for the MSSM inflation if the number of e-folds are large enough as described below. 

If the Hubble parameter of a preceding inflation is larger than $m_\phi$, the slow-roll region for the MSSM inflaton during the preceding inflation includes both the inflection point and the origin. 
$\varphi$ enters the slow-roll regime in several e-folds if it starts its evolution at a point larger than the inflection point. Furthermore after 
\begin{equation}
{\cal N}_{\rm ent}\gtrsim 10^2 \left(\frac{H}{m_\phi}\right)^2\left(\frac{M_G}{\varphi_0}\right)^2 \label{tslo}
\end{equation}
e-folds, its classical value enters the slow-roll region for the MSSM inflation \cite{Allahverdi:2008bt}. 
Therefore such a preceding inflation with a sufficiently large number of e-folds seems to naturally solve the initial value problem as long as the homogeneous field component is concerned.

\subsection{Effects of quantum fluctuations \label{PDFF}}

Here we analyze effects of quantum fluctuations on the onset of inflation using stochastic approach. 
In the stochastic approach \cite{Stoch, Starobinsky:1994bd} we divide the Heisenberg operator of the quantum field $\phi$ with some interaction potential $V(\phi)$ in the de Sitter background into a long wavelength part and a short wave length part. 
We take the cut off scale at $k = \epsilon a(t) H$, 
where $a(t)=a_0e^{Ht}$ is the scale factor and $H$ is the Hubble parameter during the prior inflation, $\epsilon$ is a small constant parameter. 
We can treat a coarse-grained or a long wavelength part, ${\bar \phi}({\bf x}, t)$, of $\phi$  as a stochastic classical field. 
In other words, we can calculate the expectation value of any function of ${\bar \phi}$, $\langle F[{\bar \phi}]\rangle$ as 
\begin{equation}
\langle F[{\bar \phi}]\rangle = \int d\varphi F(\varphi) \rho_1[\varphi({\bf x}, t)]. 
\end{equation}
Here $\rho_1$ is the one-point probability distribution function (PDF), $\rho_1[{\bar \phi}({\bf x}, t)=\varphi]\equiv \rho_1 [\varphi({\bf x}, t)]\equiv\rho_1[\varphi, t]$. 
It obeys the Fokker-Planck equation, 
\begin{equation}
\frac{\partial \rho_1 [\varphi({\bf x}, t)]}{\partial t}=\frac{1}{3H}\frac{\partial}{\partial \varphi}\{V^\prime[\varphi]\rho_1[\varphi({\bf x}, t)]\}+\frac{H^3}{8\pi^2}\frac{\partial^2 \rho_1[\varphi({\bf x}, t)]}{\partial \varphi^2} \equiv \Gamma_\varphi  \rho_1[\varphi({\bf x}, t)]. \label{FP1}
\end{equation}
Here $\varphi$ is a subdominant component of the Universe in this epoch and we identify it with the MSSM inflaton. 
The quasi-de Sitter expansion is realized not by $\varphi$ but another sector and the Hubble parameter $H$ is independent of $\varphi$. 

In general, the solution of \eqref{FP1} can be written in the form
\begin{equation}
\rho_1[\varphi, t]=\exp \left(-\frac{4\pi^2V(\varphi)}{3H^4}\right)\sum_{n=0}^\infty a_n\Phi_n(\varphi)e^{\Lambda_n(t-t_0)}, \label{PDF1}
\end{equation}
where $\Phi_n(\psi)$ is the complete orthonormal set of eigenfunctions of the equation, 
\begin{equation}
\left[ -\frac{1}{2}\frac{\partial^2}{\partial \varphi^2}+W(\varphi)\right]\Phi_n(\varphi)=\frac{4\pi^2\Lambda_n}{H^3}\Phi_n(\varphi), \label{Schr}
\end{equation}
with 
\begin{align}
W(\varphi)&=\frac{1}{2}[v^\prime(\varphi)^2-v^{\prime\prime}(\varphi)],&v(\varphi)\equiv\frac{4\pi^2}{3H^4}V(\varphi). 
\end{align}
Here the eigenvalues $\Lambda_n$'s are non-negative. 
Because $\Phi_0(\varphi)=N^{-1/2}e^{-v(\varphi)}$ is always a solution of \eqref{Schr} with the corresponding eigenvalue $\Lambda_0=0$, any solution of \eqref{PDF1} asymptotically approaches the static equilibrium solution 
\begin{equation}
\rho_{1{\rm eq}}(\varphi)=N^{-1}e^{-2v(\varphi)} \label{eq_PDF1},  
\end{equation}
with a time scale of $\Lambda_1^{-1}$ irrespective of the initial value. 
Here $N\equiv \int_{-\infty}^\infty d\varphi e^{-2v(\varphi)}$ is the normalization factor. 

In the de Sitter background with the Hubble parameter $H>m_\phi$, the slow-roll region for $\varphi$ is 
\begin{equation}
|\varphi|\lesssim \left(\frac{16}{45}\right)^{1/8}\left(\frac{H M_G^3}{\lambda}\right)^{1/4} \equiv \varphi_{\rm sr},  
\end{equation}
which includes the inflection point, $\varphi_0$.

No matter how large an initial value it takes, $\varphi$ enters the slow-roll regime after several e-folds and its PDF starts to approach the static equilibrium solution according to \eqref{PDF1}. 
Although the number of e-folds required for relaxation, $H\Lambda_1^{-1}$, is difficult to evaluate precisely due to the nontrivial shape of the potential $V[\varphi]$, we can estimate it as $H\Lambda_1^{-1} \approx {\rm max}((H/m_\phi)^2, {\cal N}_{\rm sw})$.  
Here ${\cal N}_{\rm sw}$ is the number of e-folds for quantum fluctuations of massless scalar field to sweep the plateau region $|\varphi-\varphi_0|<\varphi_0/3$ where $V[\varphi]$ is approximated by \eqref{plateau}. 
Since the quantum fluctuations of a massless field grows as $\langle \delta \varphi^2(t)\rangle^{1/2}=H{\cal N}^{1/2}/2\pi$, we find 
\begin{equation}
{\cal N}_{\rm sw}\simeq \left(\frac{4\pi}{3}\right)^2\left(\frac{\varphi_0}{H}\right)^2. \label{teqq}
\end{equation} 
We therefore find  $H\Lambda_1^{-1}\simeq {\cal N}_{\rm sw}<{\cal N}_{\rm ent}$ if
\begin{equation}
H>0.65 \left(\frac{m_\phi}{M_G}\right)\varphi_0\simeq 5.3\times 10^6 {\rm GeV}. 
\end{equation} 
In this case by the time the classical solution reaches the slow-roll region for the MSSM inflation, the dynamics of $\varphi$ would have been entirely dominated by quantum fluctuations and its PDF would have been reached to the equilibrium distribution. 
In such a situation the probability to find $\varphi$ in the slow-roll MSSM inflation region is less than $10^{-10}$. 

If $m_\phi <H<5.3\times 10^6$ GeV, on the other hand, the PDF of $\varphi$ would not have relaxed to the equilibrium state when the classical solution has reached $\varphi_{c+}$ after ${\cal N}_{\rm ent}$ e-folds. In this case the amplitude of fluctuation around the classical value is estimated as
\begin{equation}
\langle \delta \varphi(t)^2 \rangle^{1/2} \simeq \frac{H}{2\pi}{\cal N}_{\rm ent}^{1/2}\simeq \frac{5}{\pi}\frac{H^2M_G}{m \varphi_0}=1.0\times 10^7 {\rm GeV} \left(\frac{H}{m_\phi}\right)^2\left(\frac{\varphi_0}{4.0\times 10^{14}{\rm GeV}}\right)^{-1}\left(\frac{m_\phi}{10^3{\rm GeV}}\right), 
\end{equation}
which exceeds the range of the slow-roll region for the MSSM inflation. 
Then although we can find $\varphi$ to lie in the appropriate region with a finite probability, this does not guarantee that the MSSM inflation sets in because the requirement of spatial  homogeneity imposes further constraint as we see below.

\subsection{Homogeneous region in the de Sitter background}
We have seen above that the classical value of $\varphi$ approaches the slow-roll region for the MSSM inflation with sufficiently small velocity during the false vacuum inflation in some situation with a small but finite probability. 
We need, however, a homogeneous region, with $(1-10^{-10})\varphi_0<\varphi<(1+10^{-10})\varphi_0$, over a radius larger than $H_{\rm MSSM}^{-1}$.
So it is nontrivial whether required homogeneity is realized. 
In order to see if the initial condition provided by a preceding inflation satisfies the requirement, we estimate the two point correlation function  $\langle {\delta \varphi}({\bf x},t){\delta \varphi}({\bf x}+{\bf r},t) \rangle$. 
In the region where the background value of $\varphi$ is $\varphi\simeq\varphi_0$, the deviation from the classical value, $\delta \varphi({\bf x},t)\equiv \varphi({\bf x},t)-\varphi(t)$ can be approximated as a massless free field. 
In this approximation, $\langle {\delta \varphi}({\bf x},t){\delta \varphi}({\bf x}+{\bf r},t) \rangle$ reads \cite{Tagirov:1972vv}, 
\begin{align}
\langle {\delta \varphi}({\bf x},t){\delta \varphi}({\bf x}+{\bf r},t) \rangle&\simeq \langle {\delta \varphi}^2({\bf x},t) \rangle \left(1-\frac{1}{Ht}\ln HR\right)\notag \\
&=  \langle {\delta \varphi}^2({\bf x},t) \rangle-\frac{H^2}{4\pi^2}\ln HR, 
\end{align}
where $ \langle {\delta \varphi}^2({\bf x},t) \rangle =\frac{H^3}{4\pi^2}t$ and $R\equiv a_0 e^{Ht}r$ is the physical separation. 
Therefore on a scale $R\gtrsim H^{-1}$, the expectation value of the shift of the scalar field from the inflection point is about $H/2 \pi$. 
In the case the Hubble parameter of the preceding inflation is larger than the width of the slow-roll region for the MSSM inflation, $H > \varphi_0^3/(120 M_G^2)\simeq 10^4$ GeV, 
the required homogeneity for the MSSM inflation is not acquired at the end of the preceding inflation. 
Therefore it depends on the cosmic history after the end of the preceding inflation whether the MSSM inflation takes place or not.

\subsection{Condition after the preceding inflation}
In this subsection we consider the condition for the realization of the MSSM inflation after the scenarios described above. 
During the preceding inflation with the Hubble parameter $H>m_\phi$, there exists a point where $\varphi$ is set inside the slow-roll region for the MSSM inflation though its probability is rather small and inhomogeneity is somewhat large. 
Then, the Hubble parameter gradually decreases and the MSSM inflaton subsequently becomes the dominant component of the Universe when the Hubble parameter of the Universe reaches $H_{\rm MSSM}$. 

If the Hubble parameter of the preceding inflation is less than the width of the slow-roll region for the MSSM inflation, $10^4$ GeV, the spatial fluctuations do not have bad influence on the realization of the MSSM inflation. 
Note that reheating temperature must satisfy $T_R<h\varphi_0 \sim 10^{11}$ GeV to avoid high temperature correction to the potential and to suppress inhomogeneity due to thermal fluctuations.
In conclusion, the MSSM inflation can take place with a finite probability if there is not thermalization with a high reheating temperature or other mechanism that disturb the spatial homogeneity. 

If the Hubble parameter of the preceding inflation is larger than $10^4$ GeV, the required homogeneity is not satisfied at the end of the preceding inflation. 
Therefore in order for the MSSM inflation to take place, a mechanism that generates the required spatial homogeneity is needed separately. 
One possibility is another inflation with the Hubble parameter less than $10^4$ GeV. 
Because we consider the case where $\varphi$ already sets in the slow-roll region for the MSSM inflation, the number of e-folds of this secondary inflation need not be so large. 
We conclude that the MSSM inflation can take place if and only if the preceding cosmic history before the MSSM inflation satisfies the above nontrivial conditions. 

Finally we comment on the case where the number of e-folds of the MSSM inflation is smaller than $\cal{N}_{\rm COBE}$. 
In such a case, the MSSM inflaton must start slow-roll evolution near the edge of the slow-roll region, $\varphi_0-\varphi_0^3/(120M_G^2)$.  
So the required fine-tuning for the initial condition is even more stringent than discussed above.

\section{Conclusion \label{conc}}
In this paper we discussed the fine-tuning problem of the initial condition for the MSSM inflation model. 
We have considered two effects on the MSSM inflation, namely, dissipative effects due to particle creation and stochastic behavior during a prior inflation stage. 

First, we find that low-energy dissipative effects can arise on the MSSM inflation due to the interaction with other MSSM fields \cite{Moss:2006gt,Berera:2008ar,Graham:2008vu}. 
However, in the context of MSSM, the dissipative channels are restricted and the dissipative effect is too weak for the MSSM inflaton to have a thermal friction-dominated inflationary solution. 
Therefore, dissipative effects do not help to enlarge the slow-roll inflationary domain. 
Furthermore it is difficult to connect the radiation dominated Universe to the MSSM inflation because during the radiation dominated era, because there is too small number of e-folds during radiation dominated era. 
Therefore it still needs a (somewhat accidental) fine-tuning for the MSSM inflation to take place. 

Second, although it is claimed that attractor behavior in the de Sitter background can tune the initial condition of the MSSM inflaton \cite{Allahverdi:2007wh,Allahverdi:2008bt}, 
the probability that the MSSM inflaton enters the slow-roll region for the MSSM inflation becomes small due to the stochastic behavior in the quasi-de Sitter Universe, in particular,  when its number of e-folds are large enough for the PDF to approach the static equilibrium state. 
Moreover a preceding inflation with a Hubble parameter larger than $10^4$ GeV prevent the MSSM inflaton from distributing uniformly to the level required for the MSSM inflation. 
This is because stochastic force push the scalar fields about $H/(2\pi)$ after every Hubble time and it is larger than the slow-roll region. 
In this case, some additional mechanisms are needed for the MSSM inflation to take place. 
Some preceding inflation with $H <10^4$ GeV without reheating process are required for the realization of the MSSM inflation. 

\section*{Acknowledgments}

KK thanks Alexei A. Starobinsky, Fuminobu Takahashi and Ian Moss for helpful comments. 
He also thanks Daisuke Yamauchi for useful discussions. 
This work was partially supported by JSPS through research fellowships (KK) and Grant-in-Aid for Scientific Research No. 19340054(JY). This work was also supported in part by Global COE Program (Global Center of Excellence for Physical Sciences Frontier), MEXT, Japan.

\end{document}